\documentclass[superscriptaddress,aps,twocolumn]{revtex4}
\usepackage{amssymb}
\usepackage{amsmath}
\usepackage{graphicx}
\usepackage{color}
\usepackage{amsfonts}
\usepackage{multirow}

\begin{document}

\title{Terrace effects in \ grazing-incidence fast atom diffraction
from a LiF(001) surface}
\author{L. Frisco}
\affiliation{Instituto de Astronom\'{\i}a y F\'{\i}sica del Espacio (UBA-CONICET), Ciudad
Universitaria, (C1428EGA) Buenos Aires, Argentina.}
\author{M.S. Gravielle\thanks{%
Author to whom correspondence should be addressed.\newline
Electronic address: msilvia@iafe.uba.ar}}
\affiliation{Instituto de Astronom\'{\i}a y F\'{\i}sica del Espacio (UBA-CONICET), Ciudad
Universitaria, (C1428EGA) Buenos Aires, Argentina.}
\date{\today }

\begin{abstract}
The effect produced by surface defects on grazing-incidence fast atom
diffraction (GIFAD) patterns is studied by considering the presence of
terraces in a LiF(001) sample. For helium atoms impinging along the $%
\left\langle 110\right\rangle $ direction of the LiF surface, we analyze the
influence of a monolayer terrace with its edge oriented parallel or
perpendicular to the axial channel. We found that the presence of an outward
transverse step introduces a diffuse background above the Laue circle, which
displays additional peaked structures. For inward transverse steps, instead,
such a background is placed below the Laue circle, showing a much weaker
intensity. On the other hand, parallel steps give rise to asymmetric angular
distributions, which are completely confined to the Laue circle. Therefore,
these theoretical results suggest that GIFAD might be used to characterize
terrace defects.
\end{abstract}

\maketitle

\section{Introduction}

Grazing-incidence fast atom diffraction (GIFAD) is an exceptionally
sensitive technique of surface analysis, which provides detailed information
of the electronic and morphological features of the surface \cite%
{Winter2011,Debiossac2017}. During the 15 years since its first observation
\cite{Schuller2007,Rousseau2007}, the GIFAD method has been successfully
applied to study the topmost atomic layer of a wide variety of materials,
ranging from insulators \cite{Schuller2012}, semiconductors \cite%
{Debiossac2014} and metals \cite{Rios2013} to adsorbate-covered metal
surfaces \cite{Schuller2009b}, ultrathin films \cite{Seifert2010},
organic-metal interfaces \cite{Seifert2013,Momeni2018}, and graphene layers~%
\cite{Zugarramurdi2015}. But in all cases, the use of well-ordered crystal
targets was considered an important prerequisite for the observation of
interference patterns \cite{Winter2011}.

In GIFAD the periodic ordering requirement of the crystal sample is
particularly crucial along the axial direction because projectiles probe
long distances of the surface along the incidence channel, about some
hundred \text{\AA }. Hence, careful crystal manufacture and surface
preparation represent central issues in GIFAD experiments. Nevertheless,
even under extremely good cleanness conditions, real crystals have step
defects that could affect the interference patterns \cite{Farias2002}. In
the case of alkali-halide crystals, such as NaCl, KBr, and LiF, high
resolution images provided by atomic force microscopy commonly reveal the
presence of terraces or steps on the topmost layer \cite{Meyer1988,
Prohaska1994}, which are unavoidable in the preparation process of the
sample \cite{Vidal2019}.

In this paper we address how the existence of terraces in a LiF(001) surface
might affect the diffraction patterns produced by grazing impact of fast He
atoms. The He/LiF(001) system has been extensively studied with GIFAD, being
currently considered as a prototype for this phenomenon. However, all the
theoretical descriptions assume an ideal perfect crystal surface \cite%
{diaz2022}, whereas defect contributions were only qualitatively discussed
\cite{Winter2011,Debiossac2021}.

The influence of \ surface terraces on GIFAD patterns is investigated here
by considering simple crystallographic geometries: a unique up- or down-
step, oriented perpendicular or parallel to the incidence channel. Different
step locations, relative to the focus position of the atomic beam, are
analyzed. Although such geometries represent an oversimplified description
of actual LiF surfaces, they will allow us to shed light on the expected
contribution of more complex step defects.

Our study is based on the use of the surface initial value representation
(SIVR) approximation \cite{Gravielle2014,Gravielle2015}, which is a
semiquantum method that offers a satisfactory description of GIFAD in terms
of classical projectile trajectories. The projectile-surface interaction,
which is a key ingredient in all GIFAD\ simulations, is described by means
of the pairwise additive\ (PA) potential of Refs. \cite%
{Miraglia2017,Gravielle2019}. This PA potential is built as a sum of binary
interatomic potentials that represent the interaction of the atomic
projectile with individual ionic centers in the crystal, incorporating
nonlocal contributions of the electron density, along with the effect of the
Madelung potential. Concerning our theoretical model, it should be mentioned
that the combination of the SIVR approach with this PA potential has already
shown to provide GIFAD patterns in very good agreement with available
experimental data for the He/LiF(001) system under different incidence
conditions \cite{Gravielle2014,Gravielle2015,Frisco2018}. Furthermore, the
use of PA models to represent the surface interaction makes it possible to
modify the crystallographic structure to include defects, such as terraces,
without an additional computational cost, which represents an important
advantage in comparison with \textit{ab initio} calculations, like the ones
based on density functional theory.

The article is organized as follows. The theoretical model is summarized in
Sec. II, while results for step defects transverse and parallel to the axial
direction are presented and discussed in Secs. III. A and III.B
respectively. In Sec. IV we outline our conclusions. Atomic units (a.u.) are
used unless otherwise stated.

\section{Theoretical model}

Within the SIVR approximation, the effective transition amplitude for
elastic atom-surface scattering reads \cite{Frisco2018,Gravielle2018}
\begin{eqnarray}
\mathcal{A}^{\mathrm{(SIVR)}}(\mathbf{b}) &=&\int d\mathbf{R}_{o}\ f_{s}(%
\mathbf{R}_{o}-\mathbf{b})\int d\mathbf{K}_{o}\ f_{m}(\mathbf{K}_{o})  \notag
\\
&&\times a^{\mathrm{(SIVR)}}(\mathbf{R}_{o},\mathbf{K}_{o}),  \label{Aif}
\end{eqnarray}%
where
\begin{eqnarray}
a^{\mathrm{(SIVR)}}(\mathbf{R}_{o},\mathbf{K}_{o})
&=&\int\limits_{0}^{+\infty }dt\ \left\vert J_{M}(t)\right\vert
^{1/2}e^{i\nu _{t}\pi /2}\ V_{\mathrm{PS}}(\mathbf{R}_{t})  \notag \\
&&\times \exp \left[ i\left( \phi _{t}-\mathbf{Q}\cdot \mathbf{R}_{o}\right) %
\right]  \label{an0}
\end{eqnarray}%
is the partial amplitude corresponding to the classical projectile
trajectory $\mathbf{R}_{t}\equiv \mathbf{R}_{t}(\mathbf{R}_{o},\mathbf{K}%
_{o})$, which starts at the initial time $t=0$ in the position $\mathbf{R}%
_{o}$ with momentum $\mathbf{K}_{o}$. The functions $f_{s}$ $\ $and $f_{m}$
describe the spatial and momentum profiles, respectively, of the incident
projectile wave-packet, while the vector $\mathbf{b}$ denotes the initial
position of the wave-packet center. In Eq. (\ref{an0}), $J_{M}(t)=\det \left[
\partial \mathbf{R}_{t}/\partial \mathbf{K}_{o}\right] =\left\vert
J_{M}(t)\right\vert \exp (i\nu _{t}\pi )$ is the Maslov factor (a
determinant), $V_{\mathrm{PS}}(\mathbf{R}_{t})$ represents the
projectile-surface interaction along the projectile path, and $\mathbf{Q}=%
\mathbf{K}_{f}-\mathbf{K}_{i}$ is the projectile momentum transfer, with $%
\mathbf{K}_{i}$ ( $\mathbf{K}_{f}$) being the initial (final) projectile
momentum and $K_{f}=K_{i}$. The SIVR phase at time $t$ reads \cite%
{Gravielle2014}
\begin{equation}
\phi _{t}=\int\limits_{0}^{t}dt^{\prime }\ \left[ \frac{\left( \mathbf{K}%
_{f}-\mathbf{K}_{t^{\prime }}\right) ^{2}}{2m_{P}}-V_{\mathrm{PS}}(\mathbf{R}%
_{t^{\prime }})\right] ,  \label{fiSIVR}
\end{equation}%
where $m_{P}$ is the projectile mass and $\mathbf{K}_{t}=m_{P}d\mathbf{R}%
_{t}/dt$ is the classical projectile momentum.

\bigskip

In this work, the projectile-surface potential is evaluated with the PA
model of Refs. \cite{Miraglia2017,Gravielle2019}. It is expressed as

\begin{equation}
V_{\mathrm{PS}}(\mathbf{R}_{t})=\sum\limits_{\mathbf{r}_{\mathrm{B}}}v_{%
\mathbf{r}_{\mathrm{B}}}\left( \mathbf{R}_{t}-\mathbf{r}_{\mathrm{B}}\right)
+U_{\mathrm{PS}}(\mathbf{R}_{_{t}}),  \label{VPS}
\end{equation}%
where $v_{\mathbf{r}_{_{\mathrm{B}}}}\left( \mathbf{r}\right) $ describes
the short-range binary interaction between the projectile and the crystal
ion placed at the Bravais lattice site $\mathbf{r}_{_{\mathrm{B}}}$, as a
function of the relative vector $\mathbf{r}$, and $U_{\mathrm{PS}}(\mathbf{R}%
_{_{t}})$ denotes the projectile polarization term, which describes the
long-range projectile-surface interaction due to the rearrangement of the
electron density of the atomic projectile.

The binary potentials $v_{\mathbf{r}_{_{\mathrm{B}}}}$ are expressed in
terms of the unperturbed electron densities of the projectile and the ionic
sites, incorporating not only non-local contributions of these electron
densities, but also the Madelung contribution, i.e., the effect of the ionic
crystal lattice on the electron density around each individual ionic site.
In turn, the potential $U_{\mathrm{PS}}$ depends on the surface electric
field at the position $\mathbf{R}_{_{t}}$ of the atomic projectile, reading
\begin{equation}
U_{\mathrm{PS}}(\mathbf{R}_{_{t}})=-\frac{\alpha _{P}}{2}\left\vert
\sum\limits_{\mathbf{r}_{\mathrm{B}}}\mathbf{E}_{\mathbf{r}_{\mathrm{B}}}(%
\mathbf{R}_{t}-\mathbf{r}_{\mathrm{B}})\right\vert ^{2},  \label{Upol}
\end{equation}%
where $\alpha _{P}$ is the dipole polarizability of the projectile ($\alpha
_{P}=1.38$ a.u. for He atoms) and $\mathbf{E}_{\mathbf{r}_{\mathrm{B}%
}}\left( \mathbf{r}\right) $ is the electric field produced by the
asymptotic charge of the crystal ion placed at $\mathbf{r}_{_{\mathrm{B}}}$.
In Eqs. (\ref{VPS}) and (\ref{Upol}) the summation on $\mathbf{r}_{\mathrm{B}%
}$ covers all the occupied lattice sites of the crystal sample, taking into
consideration the presence of flat terraces, as well as the rumpling of the
topmost atomic layer of each terrace, while the sub-index $\mathbf{r}_{%
\mathrm{B}}$ in $v_{\mathbf{r}_{\mathrm{B}}}$ and $\mathbf{E}_{\mathbf{r}_{%
\mathrm{B}}}$ was included to distinguish the the two different ions of the
crystallographic basis.

From Eq. (\ref{Aif}), the differential probability of scattering in the
direction of the solid angle $\Omega _{f}=(\theta _{f},\varphi _{f})$, with $%
\theta _{f}$ ($\varphi _{f}$) being the final polar (azimuthal) angle
measured with respect to the surface (axial channel), can be obtained as:
\begin{equation}
\frac{dP^{\mathrm{(SIVR)}}}{d\Omega _{f}}=K_{f}^{2}\int d\mathbf{b\ }%
\left\vert \mathcal{A}^{\mathrm{(SIVR)}}(\mathbf{b})\right\vert ^{2},
\label{dP0}
\end{equation}%
where the\ integral on $\mathbf{b}$ covers an area equal to a reduced unit
cell of the crystal surface. This integration, associated with the spot-beam
effect \cite{Frisco2018}, takes into account that it is experimentally
impossible to control the focus position of the atomic beam with nanoscale
precision. Details about the SIVR method can be found in Refs. \cite%
{Gravielle2014,Gravielle2015,Frisco2018,Gravielle2018}.

\section{Results}

\begin{figure}[tbp]
\includegraphics[width=0.4 \textwidth]{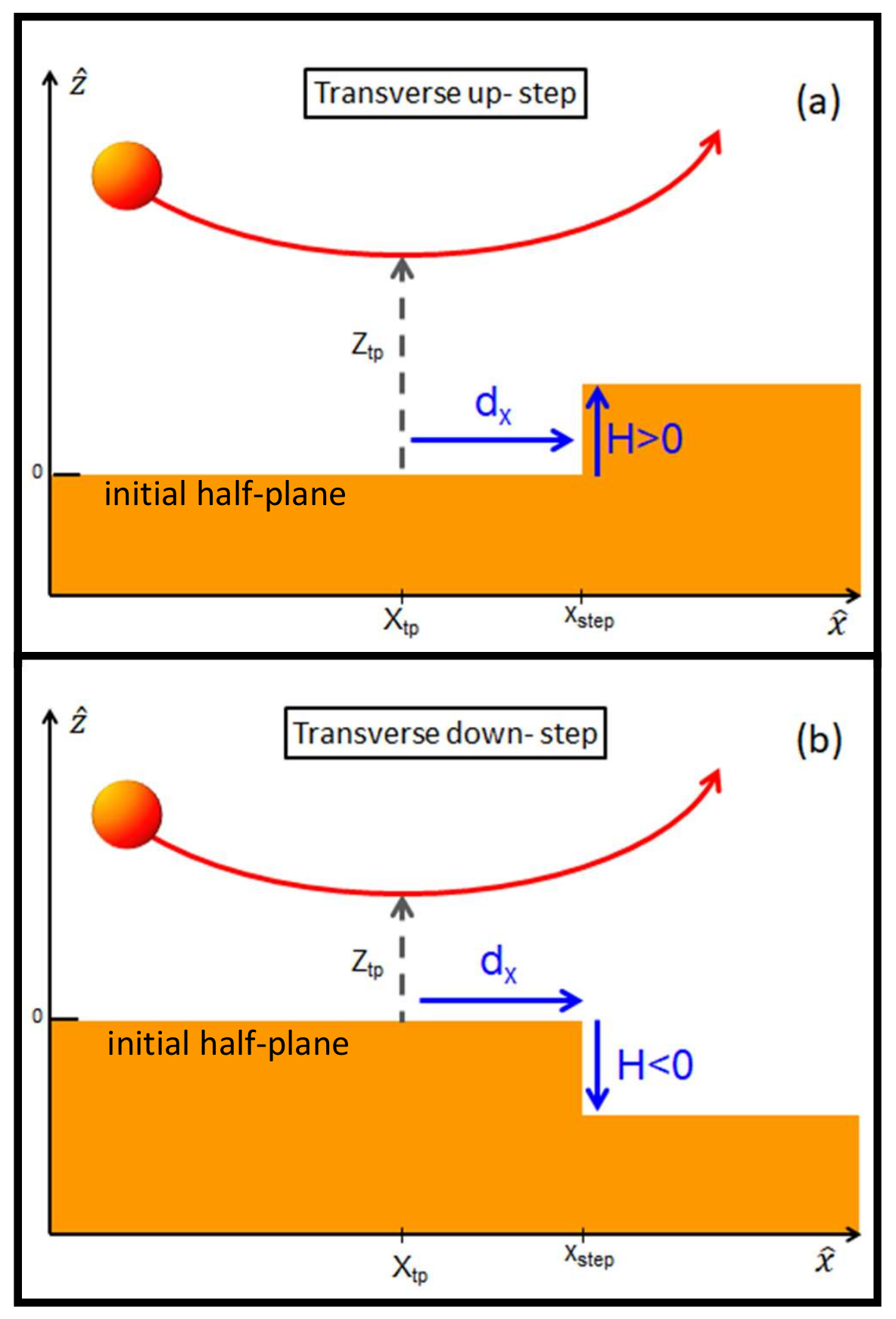} \centering
\caption{(Color online) Depiction of \ (a) up- and (b) down- steps oriented
perpendicular to the axial channel ($\widehat{x}$).}
\label{esquema}
\end{figure}

With the aim of studying the effect produced by the presence of terraces in
alkali-halide crystal surfaces, we evaluated diffraction patterns for $^{4}$%
He atoms scattered off \ LiF(001) along the $\left\langle 110\right\rangle $
direction considering a crystal sample with\ one step of height $H$ oriented
perpendicular or parallel to the incidence channel. The height of the step
was assumed to be equal to the distance between layers, that is, $H=\pm a/2$%
, where $a$ is the lattice parameter ($a=4.02$ \AA\ for LiF) and the sign $%
\pm $ indicates an outward ($+$) or inward ($-$) terrace parallel to the
surface plane (see Fig. \ref{esquema}). Notice that for alkali-halide
surfaces, atomically flat terraces with sizes ranging from $1000$ to $2000$
\AA\ can be observed under the usual cleanness conditions \cite{Prohaska1994}%
. But the heights of these terraces are also variable and they can be higher
than the monolayer height \cite{Meyer1988}.

The terrace was simulated by adding or removing a monolayer (according to
the sign of $H$) in the topmost half-plane of the crystal sample, as
depicted in Fig. \ref{esquema}. \ The topmost atomic layer of the terrace
includes the rumpling, that is, the different relaxation of the outermost F$%
^{-}$\ and Li$^{+}$ ions with respect to the unreconstructed plane \cite%
{Miraglia2017}. However, our simplified step model neglects the
reconstruction effects at the edge of the terrace, describing the crystal
defect as a sharp step. Even though such an ideal cutting of the crystal
terrace represents a rough description of real surface defects, this
assumption can be consider as a zero-order approach to study the influence
of step defects in GIFAD. Moreover, recent topographic images of other
alkali-halide crystal - KBr - show sharp profiles of the monatomic steps
\cite{Ito2016}.

\bigskip

In this article, we chosen a fixed impact energy, $E=$ $%
K_{i}^{2}/(2m_{P})=1.25$ keV, for which experimental GIFAD distributions are
available \cite{Schuller2009c}. Two normal incidence energies $E_{\perp
}=E\sin ^{2}\theta _{i}$ ($\theta _{i}$ denotes the glancing incidence
angle) are analyzed - $E_{\perp }=0.20$ and $0.46$ eV - which correspond
respectively to low and intermediate $E_{\perp }$ values for He/LiF GIFAD
\cite{Schuller2009}. \ Notice that for the latter normal energy, the\
experimental projectile distribution reported in Ref. \cite{Schuller2009c}
was adequately described with our theoretical method without taking into
account the presence of crystal defects \cite{Frisco2018}. Then, such a good
theory-experiment agreement suggests that typical GIFAD experiments are
carried out on large flat terraces, being unaffected by step defects \cite%
{Debiossac2021}.

For different configurations and relative positions of the monolayer step,
two-dimensional projectile distributions, as a function of the final polar
and azimuthal angles, were calculated within the SIVR\ approach, as given by
Eq. (\ref{dP0}). In the calculation of $\mathcal{A}^{\mathrm{(SIVR)}}$ [Eq. (%
\ref{Aif})], the spatial and momentum profiles of the incident wave packet \
were determined as given in Refs. \cite{Gravielle2015,Gravielle2018} by
considering a collimating scheme formed by a rectangular slit of area $0.40$
$\times 0.09$ mm$^{2}$ (the latter length in the transverse direction)
placed at $30$ cm from the surface, with an angular beam dispersion of $%
0.006 $ deg. From these collimating parameters, which are in agreement with
current experimental setups \cite{Bocan2020}, we derived the transverse
coherence lengths in the directions perpendicular ($\sigma _{y}$) and
parallel ($\sigma _{x}$) to the incidence channel \cite%
{Gravielle2015,Gravielle2018} - $\sigma _{y}\simeq 13$ \AA\ and $\sigma
_{x}\simeq 240$ ($160$) \AA\ for $E_{\perp }=0.20$ ($0.46$) eV - which
verify the relation $\sigma _{x}\gg \sigma _{y}$. It reinforces the fact
that under grazing incidence, helium projectiles probe much longer distances
along the axial channel than in the perpendicular direction, making GIFAD
patterns more sensitive to transverse steps, oriented perpendicular to the
channel, than to those oriented in the parallel direction.

\subsection{Effects due to a transverse terrace}

In this subsection, we consider a LiF(001) surface with a monolayer step
oriented perpendicular to the axial direction. To investigate the influence
of the position of the border of the terrace on the diffraction patterns, we
determine the relative position of the transverse step by means of the
distance $d_{x}$ between the edge of the terrace and the mean focus point of
the incident beam, that is,

\begin{equation}
d_{x}=x_{\mathrm{step}}-X_{\mathrm{F}},  \label{distance}
\end{equation}%
where $x_{\mathrm{step}}$ is the step position along the incidence channel ($%
\widehat{x}$)\ and $X_{\mathrm{F}}$ denotes $\ $the mean position along $%
\widehat{x}$ of the focus point of the helium beam. The value of $X_{\mathrm{%
F}}$ is defined as the average of the $x$- positions corresponding to the
turning points $(X_{tp},Z_{tp})$ of projectile trajectories specularly
reflected from a perfect surface (see Fig. \ref{esquema}). For an ideal%
\textit{\ }LiF(001) crystal, these trajectories run on the flat regions of
the projectile-surface potential, i.e., along the F$^{-}$ or Li$^{+}$ rows.

Using the relative step position defined by Eq. (\ref{distance}), positive
distances $d_{x}$ indicate that He projectiles should reach the turning
point before being affected by the step in the outgoing path, while negative
values are associated with steps affecting the incoming path of the incident
atoms. Clearly, this is only an overall description of the scattering
process in the presence of surface terraces because depending on the
incidence conditions and the $d_{x}$ value, the turning points of scattered
projectiles could be modified by the strong change in the surface potential
introduced by the presence of the outward or inward step.

\bigskip

We start analyzing projectile distributions for the higher normal energy - $%
E_{\perp }=0.46$ eV - for which terrace effects are expected to be more
important. For this normal energy, corresponding to the incidence angle $%
\theta _{i}=1.1\deg $, simulated diffraction patterns respectively derived
from a perfect LiF(001) surface and from a surface with an upward step
placed at a distance $d_{x}\simeq +120$ \AA\ are compared in Fig. \ref%
{e046-perf-x700up}. Taking into account that large terraces are usually
present in the LiF samples used in GIFAD experiments \cite{Debiossac2021},
this distance $d_{x}$ corresponds to a step position relatively close to the
beam focus point.

\begin{figure}[tbp]
\includegraphics[width=0.5 \textwidth]{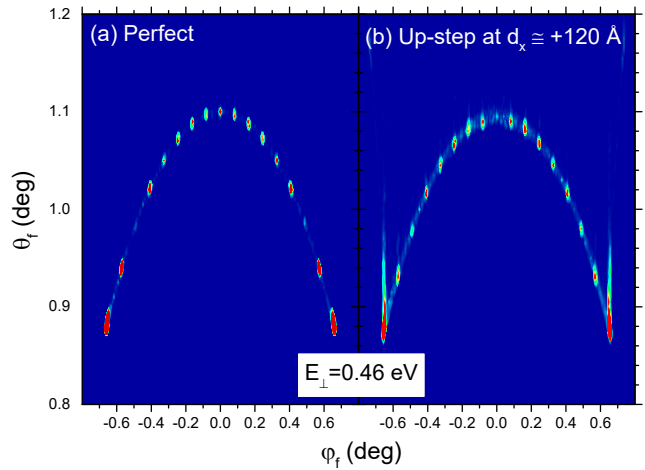} \centering
\caption{(Color online) Two-dimensional projectile distributions, as a
function of $\protect\theta _{f}$ and $\protect\varphi _{f}$, for $1.25$ keV
$^{4}$He atoms scattered off LiF(001) along the $\left\langle
110\right\rangle $ channel with $E_{\perp }=0.46$ eV. SIVR simulations for
(a) a perfect crystal surface and (b) a surface with a transverse up-step
placed at $d_{x}\simeq +120$ \AA , with $d_{x}$ as defined in Eq. (\protect
\ref{distance}), are displayed.}
\label{e046-perf-x700up}
\end{figure}

Both panels of Fig. \ref{e046-perf-x700up} present similar interference
structures, with equally $\varphi _{f}$- spaced Bragg maxima lying on a thin
annulus associated with the Laue circle \cite{Rousseau2007,Schuller2008},
which is defined by $\theta _{f}^{(\text{\textrm{L}})}=\left( \theta
_{i}^{2}-\varphi _{f}^{2}\right) ^{1/2}$. However, as a consequence of the
presence of the upward step, the angular distribution of Fig. \ref%
{e046-perf-x700up} (b) shows outermost peaks noticeably extended towards
larger polar angles, along with different relative intensities of the inner
peaks, in comparison with those for a perfect surface (Fig. \ref%
{e046-perf-x700up} (a)). \ Regarding this latter effect, notice that the
intensities of the Bragg maxima are modulated by the intra-channel
interference, associated with the profile of the surface potential across
the axial direction \cite{Schuller2008,Gravielle2014}, which suffers a local
change at the edge of the terrace. Instead, the $\varphi _{f}$ positions of
the Bragg peaks are determined by the inter-channel interference, which
depends on the spacing between equivalent parallel channels \cite%
{Schuller2008,Gravielle2014}, a parameter that is not altered by the
transverse step.

\begin{figure}[tbp]
\includegraphics[width=0.4 \textwidth] {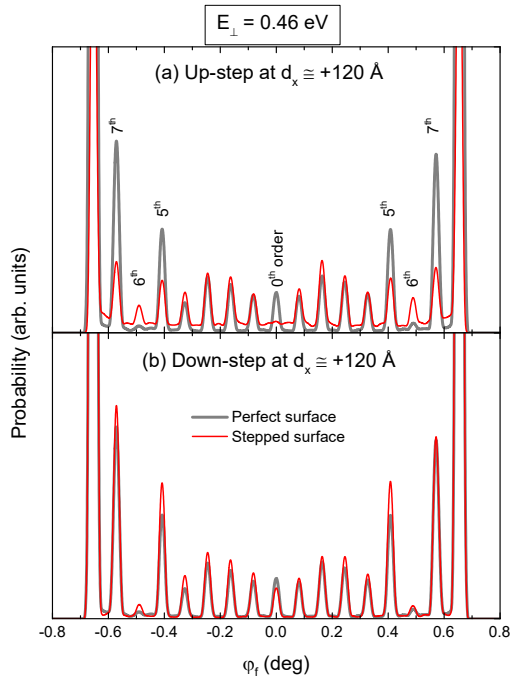} \centering
\caption{(Color online) Azimuthal spectra, corresponding to the Laue
annulus, for the normal energy $E_{\perp }=0.46$ eV. Thin red line, SIVR
simulations for a transverse (a) up-step and (b) down-step, both placed at $%
d_x \simeq +120$ \AA ; thick gray line, SIVR results for a perfect crystal
surface.}
\label{e046-x700spectra}
\end{figure}

To thoroughly analyze the step effect on Bragg-peak intensities, in Fig. \ref%
{e046-x700spectra} (a) we compare the azimuthally projected distributions
corresponding to the two panels of Fig. \ref{e046-perf-x700up}, obtained by
integrating Eq. (\ref{dP0}) over a reduced area on the Laue circle \cite%
{Debiossac2016}. In Fig. \ref{e046-x700spectra} (a) we observe that the
presence of the transverse up-step introduces an almost constant background
in the azimuthal spectrum. Furthermore, the distortion of the surface
potential caused by the outward step strongly affects the peak intensities,
markedly reducing the relative intensity of the maxima of orders $5^{th}$
and $7^{th}$, as well as that of the central peak ($0^{th}$ order) whose
intensity becomes much lower than those of the adjacent maxima, in contrast
to what is observed for a perfect surface. Also, the intensity of the
outermost maxima (not shown in Fig. \ref{e046-x700spectra} (a) ) is lowered
as a result of the surface defect, whereas the intensity of $6^{th}$- order
maxima is raised, making these latter peaks visible.

\begin{figure}[tbp]
\includegraphics[width=0.5 \textwidth] {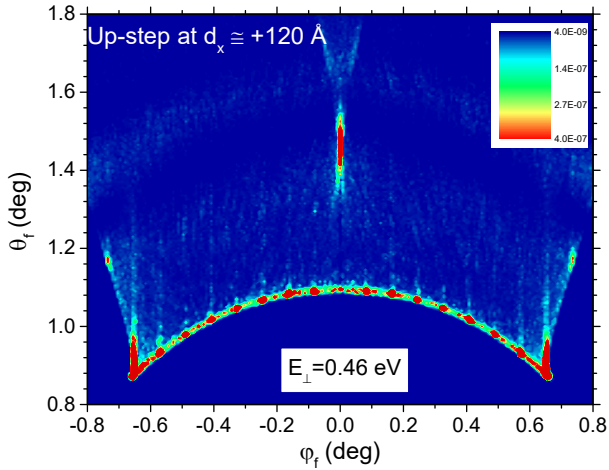} \centering
\caption{(Color online) Analogous to Fig. \protect\ref{e046-perf-x700up}
(b), extending the $\protect\theta _{f}$ range and increasing the
sensitivity of the intensity scale, as explained in the text. }
\label{e046-x700up-ext}
\end{figure}

At this point, it is important to recall that typical GIFAD patterns for
perfect crystal surfaces are essentially determined by the surface potential
averaged along the axial direction \cite{Zugarramurdi2012,Muzas2016}. But
the presence of a transverse step breaks the translation invariance
associated with this averaged surface potential, altering the \textit{%
effective }slope of the reflection plane around the edge of the terrace.
This fact is evidenced in Fig. \ref{e046-x700up-ext}, where we have extended
the $\theta _{f}$ range of Fig. \ref{e046-perf-x700up} (b), lowering also
the intensity scale by one order of magnitude to show the terrace effects.
The projectile distribution of \ Fig. \ref{e046-x700up-ext} presents a
diffuse background for $\theta _{f}\succsim \theta _{f}^{(\text{\textrm{L}}%
)} $, with additional peaks at the central and outermost azimuthal angles.
In this case, a large proportion of the scattered projectiles ($\approx 54\%$
) are deflected above the Laue circle due to the steep increase of the
surface potential at the step position, while about\ $15\%$ of the incident
projectiles penetrates in the terrace bulk.

\begin{figure}[tbp]
\includegraphics[width=0.5 \textwidth] {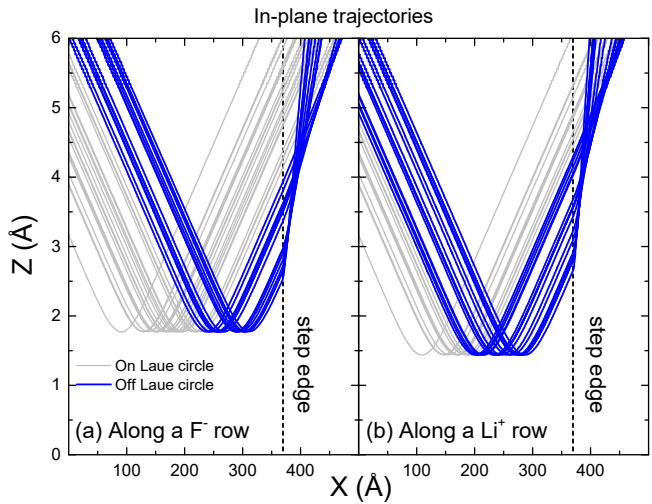} \centering
\caption{(Color online) Sample of projectile trajectories contributing to
the central region (i.e., in-plane trajectories) of Fig. \protect\ref%
{e046-x700up-ext}, after running along (a) F$^{-}$ and (b) Li$^{+}$ rows of
the initial half-plane (see Fig. \protect\ref{esquema}). Both panels display
the normal position $Z$ as a function of the coordinate $X$ along the $%
\langle 110\rangle $ channel. Gray and blue lines indicate trajectories
ending respectively \textit{on} and \textit{off} the Laue circle.}
\label{e046-x700up-tray}
\end{figure}

Since the most remarkable feature of the projectile distribution of Fig. \ref%
{e046-x700up-ext} is the central peak above the Laue circle, which is placed
at $\theta _{f}\simeq 1.47\deg $, in Fig. \ref{e046-x700up-tray} we show a
sample of random projectile paths that contribute to the central region of
this angular distribution. Such trajectories are confined to the scattering
plane (i.e., in-plane trajectories), initially running along the F$^{-}$ or
Li$^{+}$ rows of the topmost layer in the initial half-plane (note that F$%
^{-}$ or Li$^{+}$ rows switch at the terrace), as respectively shown in
Figs. \ref{e046-x700up-tray} (a) and \ref{e046-x700up-tray} (b). \ In both
panels, the wide $x$- spread of the classical turning points is associated
with the large $\sigma _{x}$ value. Hence, projectile paths with turning
points far away from the step end on the Laue circle, without being affected
by the presence of the upward terrace. But those trajectories with turning
points placed at distances $\left\vert x_{\mathrm{step}}-X_{tp}\right\vert
\lesssim 130$ ($170$) \AA\ in Fig. \ref{e046-x700up-tray} (a)\ (Fig. \ref%
{e046-x700up-tray} (b)), are deflected with a final polar angle $\theta
_{f}\succ \theta _{f}^{(\text{\textrm{L}})}$ due to the change in the
surface potential, which abruptly becomes repulsive at the step. We stress
that in Fig. \ref{e046-x700up-ext}, the \textit{on}-Laue interference
maxima, as well as the intense outermost peaks above the Laue circle, are
produced by quantum interference among partial transition amplitudes
corresponding to different projectile paths, that is, they cannot be
explained as points of accumulation of trajectories. However, the central
maximum above the Laue circle has a classical origin, also being observed in
the classical projectile distribution.
\begin{figure}[tbp]
\includegraphics[width=0.5 \textwidth] {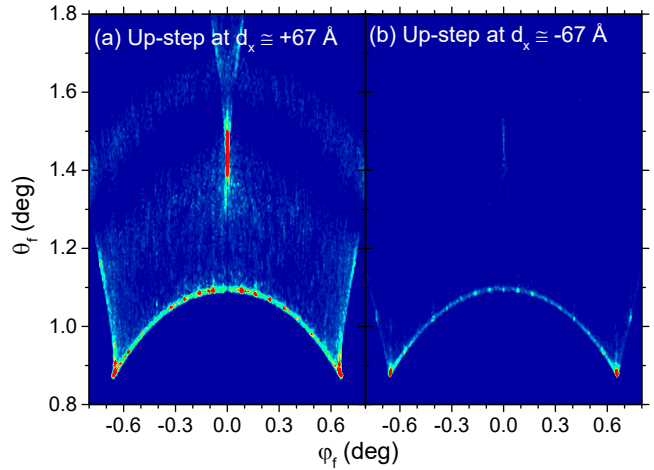} \centering
\caption{(Color online) Analogous to Fig. {\protect\ref{e046-perf-x700up}}
(b) for a LiF surface with a transverse up-step placed at (a) $d_{x}\simeq
+67$ \AA\ and (b) $d_{x}\simeq -67$ \AA .}
\label{e046-x350-600up}
\end{figure}

A similar central \textit{off}-Laue maximum, associated with the presence of
the transverse terrace, can be seen in Fig. \ref{e046-x350-600up} (a) for $%
d_{x}\simeq +67$ \AA , a step position even closer to $X_{\mathrm{F}}$. \
But in this case, since the atomic beam is almost focused on the step
region, Bragg peaks start to wash out, whereas the central peak above the
Laue circle is clearly visible without increasing the sensitivity of the
intensity scale. The same up-step effects are also observed \ in the
projectile distribution of Fig. \ref{e046-x350-600up} (b) corresponding to
an upward step placed in front of the focus point, at $d_{x}\simeq -67$ \AA %
. Nevertheless, the intensity of the background and the \textit{off}-Laue
peaks markedly decreases as the projectile trajectories \ meet the terrace
edge in their incoming paths.

\begin{figure}[tbp]
\includegraphics[width=0.5 \textwidth] {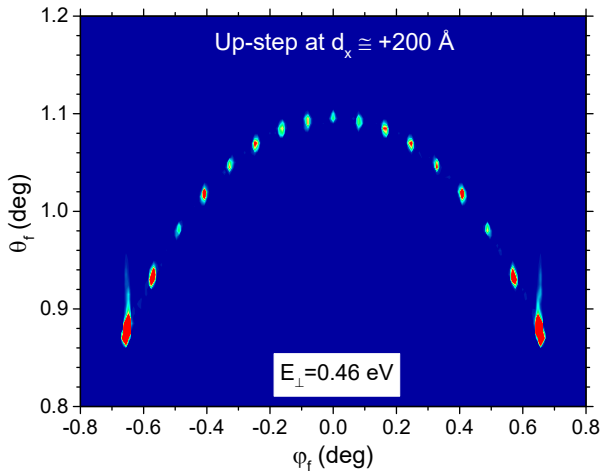} \centering
\caption{(Color online) Analogous to Fig. \protect\ref{e046-perf-x700up} (b)
for a LiF surface with a transverse up-step placed at $d_{x}\simeq +200$ \AA %
.}
\label{e046-x850up}
\end{figure}

As expected, the step effects gradually disappear as the distance $d_{x}$
increases. The projectile distribution for an upward step placed at $%
d_{x}\simeq +200$ \AA , plotted in Fig. \ref{e046-x850up}, looks similar to
that for a perfect crystal surface (Fig. \ref{e046-perf-x700up} (a)),
without any signature of \ the terrace effects, except for the polar
elongation of the outermost peaks with respect to those derived from a
perfect surface. Under these incidence conditions, most helium projectiles ($%
\approx 70\%$) hit the detector plane with $\theta _{f}\simeq \theta _{f}^{(%
\text{\textrm{L}})}$, while only a very small fraction ($\approx 2\%$)
penetrates into the bulk at the step, indicating that the projectile
distribution tends to the one corresponding to a perfect LiF surface.

\begin{figure}[tbp]
\includegraphics[width=0.5 \textwidth] {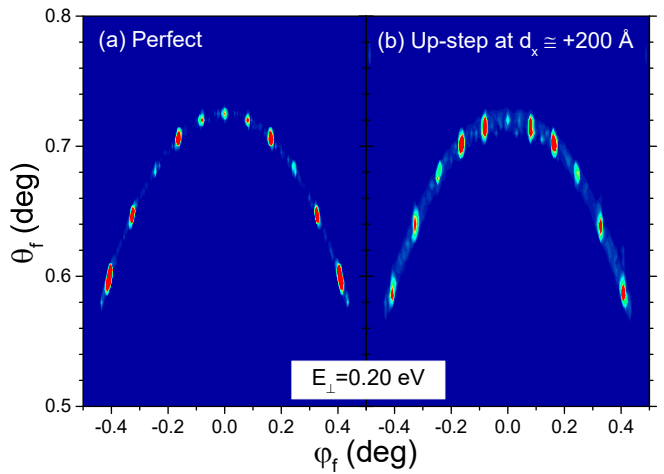} \centering
\caption{(Color online) Analogous to Fig. \protect\ref{e046-perf-x700up} for
the normal energy $E_{\perp }=0.20$ eV and a transverse up-step placed at $%
d_{x}\simeq +200$ \AA .}
\label{e020-perf-x1050up}
\end{figure}

The distance $d_{x}$ for which unperturbed GIFAD patterns can be obtained
depends on the normal energy. When $\theta _{i}$ decreases, so $E_{\perp }$
does, grazing projectiles probe longer distances along the axial direction,
being affected by transverse upward steps placed at longer distances $d_{x}$%
. This fact can be observed in Fig. \ref{e020-perf-x1050up} (b), where the
projectile distribution for $E_{\perp }=0.20$ eV ( $\theta _{i}=0.7\deg $),
produced by a crystallographic configuration similar to that of Fig. \ref%
{e046-x850up}, shows noticeable changes in the relative peak intensities,
with respect to those derived by considering a perfect crystal surface,
displayed in Fig. \ref{e020-perf-x1050up} (a).

\begin{figure}[tbp]
\includegraphics[width=0.5 \textwidth] {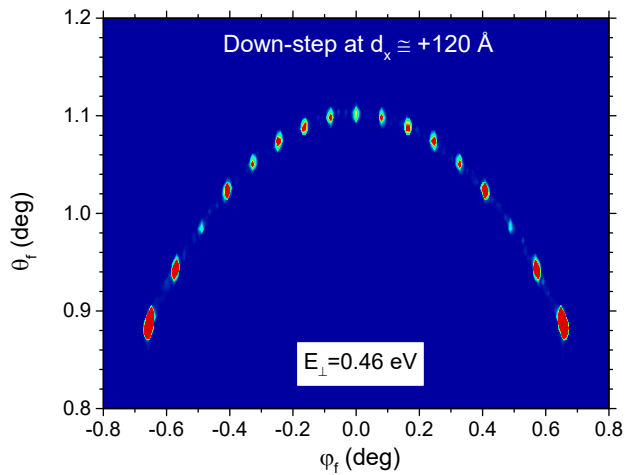} \centering
\caption{(Color online) Analogous to Fig. \protect\ref{e046-perf-x700up} (b)
for a LiF surface with a transverse down-step placed at $d_{x}\simeq +120$
\AA .}
\label{e046-x700down}
\end{figure}

Finally, we analyze the influence of a transverse down-step, as depicted in
Fig. \ref{esquema} (b). In Fig. \ref{e046-x700down}, we plot the angular
distribution for $E_{\perp }=0.46$ eV derived by considering an inward step
placed at\ $d_{x}\simeq +120$ \AA . In contrast to the up-step effects shown
in Fig. \ref{e046-perf-x700up} (b), the presence of the down step does not
affect the \textit{on}-Laue distribution, which is similar to that for a
perfect surface (Fig. \ref{e046-perf-x700up} (a)). This feature can be
quantitative confirmed by comparing the corresponding azimuthally projected
spectra, which are quite alike, as observed in Fig. \ref{e046-x700spectra}
(b). In the case of Fig. \ref{e046-x700down}, however, although
approximately $80\%$ of the trajectories are scattered from the LiF surface
with $\theta _{f}\simeq \theta _{f}^{(\text{\textrm{L}})}$, there are
trajectories whose outgoing paths are distorted by the decrease of the
potential at the edge of the terrace, being deflected below the Laue circle.
It gives rise to a diffuse background below the Laue circle, whose intensity
is several orders of magnitude lower than those of Bragg peaks.

\subsection{Effects due to a parallel terrace}

The terrace effects described in Sec. III. A change when the monolayer step
is oriented parallel to the axial direction. Analogously to the case of
transverse terraces, we determine the relative position of the parallel step
by means of the distance $d_{y}$ between the edge of the terrace and the
mean focus point, that is,

\begin{equation}
d_{y}=y_{\mathrm{step}}-Y_{\mathrm{F}},  \label{dy}
\end{equation}%
where $y_{\mathrm{step}}$ is the step position across the incidence channel (%
$\widehat{y}$)\ and $Y_{\mathrm{F}}$ denotes $\ $the mean position of the
focus point of the incident beam along $\widehat{y}$. As a consequence of
the symmetry of the problem, only outward steps placed at positive distances
$d_{y}$ will be considered in this subsection.
\begin{figure}[tbp]
\includegraphics[width=0.5 \textwidth] {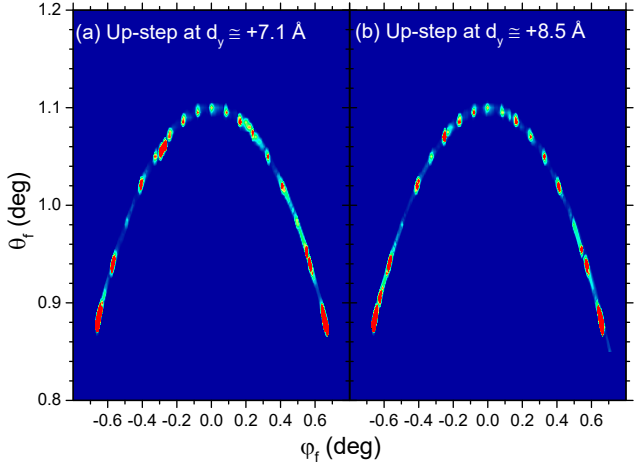} \centering
\caption{(Color online) Analogous to Fig. \protect\ref{e046-perf-x700up} (b)
for a LiF surface with a parallel up-step placed along (a) a Li$^{+}$ row
(at $d_{y}=+7.1$ \AA ), and (b) a F$^{-}$ row (at $d_{y}=+8.5$ \AA ) of the
perfect crystal surface.}
\label{e046-y}
\end{figure}

\begin{figure}[tbp]
\includegraphics[width=0.5 \textwidth] {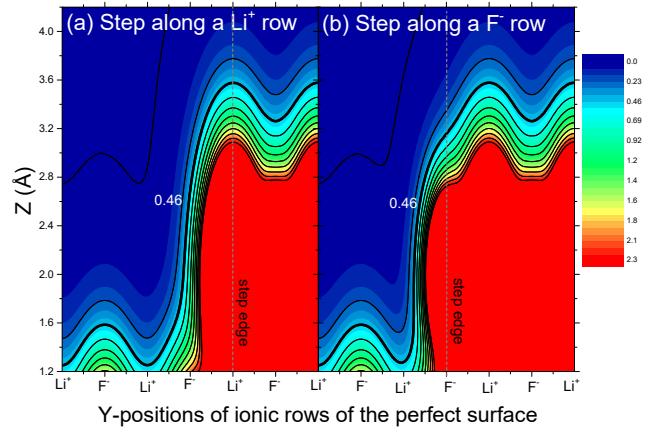} \centering
\caption{(Color online) Equipotential curves of the axial potential
(averaged along the $\langle 110\rangle $ channel) for a LiF(001) surface
with a parallel up-step placed along (a) a Li$^{+}$ row and (b) a F$^{-}$
row of the perfect crystal surface. Energy scale in eV.}
\label{poten-y}
\end{figure}

Taking into consideration that the transverse length of the surface area
that is probed by helium projectiles is much smaller than the axial one, in
Fig. \ref{e046-y} we show two-dimensional angular distributions for surface
steps along the $\left\langle 110\right\rangle $ channel considering closer
distances to the focus point of the beam, that is, $d_{y}\simeq +7.1$ \AA\ %
in Fig. \ref{e046-y} \ (a) and $d_{y}\simeq +8.5$ \AA\ in Fig. \ref{e046-y}
\ (b). These steps are placed respectively on top of Li$^{+}$ and F$^{-}$
rows of the ideal perfect surface. We found that the angular distributions
of Fig. \ref{e046-y} are fully confined to the Laue circle, that is, all
scattered projectiles leave the surface with $\theta _{f}\simeq \theta
_{f}^{(\text{\textrm{L}})}$, while the fraction of trajectories penetrating
into the terrace bulk is lower than $1\%$. Nonetheless, the presence of a
parallel step in the area that is coherently illuminated by the atomic beam
\ introduces an azimuthal asymmetry in the GIFAD patterns of Figs. \ref%
{e046-y} \ (a) and \ref{e046-y} \ (b), which display some interference
maxima \ with structures elongated along the Laue circle. This latter effect
depends on the exact position of the step, which determines the shape of the
equipotential curves in the region of the terrace edge. In Fig. \ref{poten-y}
\ we plot the equipotential curves, corresponding to surface potential
averaged along the axial direction, for the two cases of Fig. \ref{e046-y}.
Around the step region, the equipotential contours vary if the terrace edge
is along a Li$^{+}$ or a F$^{-}$ row of the perfect surface. Similar
differences in the equipotential curves are also observed for transverse
terraces. But the influence of such border effects on GIFAD patterns seems
to be stronger for parallel than for transverse upward steps.

\begin{figure}[tbp]
\includegraphics[width=0.4 \textwidth] {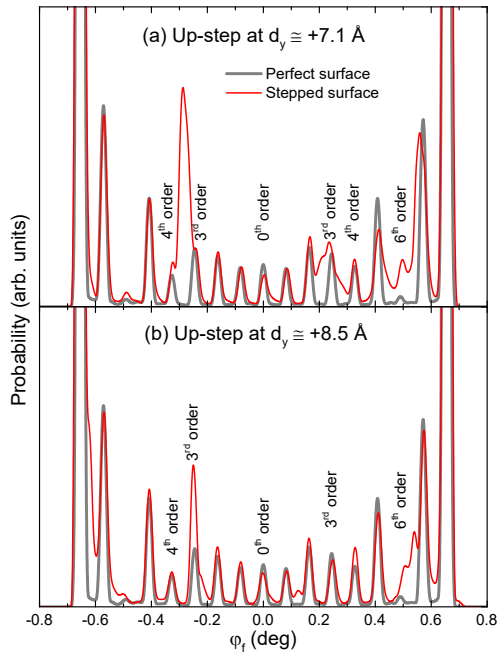} \centering
\caption{(Color online) Azimuthal spectra, corresponding to the Laue
annulus, for the cases of Fig. \protect\ref{e046-y}. Lines, analogous to
Fig. \protect\ref{e046-x700spectra}.}
\label{e046-y-spectra}
\end{figure}

To investigate in more detail the asymmetry introduced by parallel steps, in
Fig. \ref{e046-y-spectra} we plot the azimuthal spectra corresponding to the
distributions of Fig. \ref{e046-y}, contrasting them with that derived from
a perfect crystal surface. In both panels of Fig. \ref{e046-y-spectra}, the
intensities of the Bragg peaks of order $-3$ and $+6$ increase as a
consequence of the presence of the parallel step, while the central region
of the spectrum is not affected by the surface defect. Furthermore, in Fig. %
\ref{e046-y-spectra} (a) the sharp increase of the $3^{rd}$-order maxima
smudges the $4^{th}$-order peak, both forming a broad intense peak on the
left side of the projectile distribution, whereas in Fig. \ref%
{e046-y-spectra} (b), the peak of order $-4$ is not altered by the terrace
border.

\section{Conclusions}

In this work, the influence of surface defects on GIFAD patterns for the
He/LiF(001) system was investigated by considering the presence of a
monolayer step in the coherently illuminated region of the crystal target.
Our theoretical description was based on the use the SIVR method to
represent the grazing scattering process, combined with a PA atom-surface
potential \cite{Miraglia2017}. This potential model allowed us to easily
incorporate outward or inward monolayer terraces by adding or removing,
respectively, an atomic layer in a portion of the LiF sample.

In order to simplify the \ analysis, only two different orientations of the
terrace edge - transverse and parallel to the incidence channel - were
considered. For transverse up-steps placed close to the focus point of the
incident beam, at distances $d_{x}$ smaller than a few hundred \AA ,
simulated projectile distributions display the characteristic Bragg peaks on
the Laue circle, along with a diffuse background above the Laue circle,
which presents additional central and outermost peaks. The upward step also
affects the relative intensities of the \textit{on}-Laue maxima, which are
relevant for the use of GIFAD as a surface analysis technique. For
transverse down-steps, instead, the projectile distribution approximates the
one derived from a perfect crystal surface, while the terrace border
introduces only a very weak background below the Laue circle, whose
intensity is several orders of magnitude smaller than those of the \textit{on%
}-Laue maxima. As expected, these terrace effects depend on the incidence
conditions, gradually disappearing as $d_{x}$ increases.

On the other hand, as a consequence of the smaller transverse coherence
length of the beam in the direction perpendicular to the axial channel, the
presence of a parallel step affects the projectile distribution only if the
step is placed at a distance $d_{y}$ less than a few tens \AA\ from the
focus point of the incident beam. In this case, the parallel up-step
introduces an azimuthal asymmetry in the angular spectrum, which is fully
localized on the Laue circle.

Summarizing, we found that terrace effects on GIFAD patterns strongly depend
on the orientation of the edge of the monolayer terrace, as well as on the
height (outward or inward) and the position of the step. Therefore, these
findings suggest that GIFAD may be a useful tool to study terrace defects on
alkali-halide surfaces. We hope that this study will be helpful to trigger
experimental research on this topic.

\begin{acknowledgments}
M.S.G is grateful to H. Khemliche for the helpful discussions. We
acknowledge funding from ANPCYT (PICT-2017-1201; PICT-2017-2945;
PICT-2020-1755; PICT-2020-1434) of Argentina.
\end{acknowledgments}

\bibliographystyle{unsrt}
\bibliography{HeLiF-terraces}

\end{document}